# Vibration analysis of a pre-stressed graphene sheet embedded in a deformable matrix

K B Mustapha[1*]

School of Mechanical and Aerospace Engineering, Nanyang Technological University, 50 Nanyang Avenue, Singapore 639798, Republic of Singapore

**Abstract**

The effect of the initial uniaxial stress and a surrounding elastic matrix on the transverse vibration response of a single-layered graphene sheet (SLGS) is investigated through a theoretical formulation that is based on the nonlocal Kirchhoff plate theory. The surrounding elastic matrix of the SLGS is modeled as a foundation of the Pasternak-type. The elliptic partial differential equation governing the dynamics of the pre-stressed SLGS is solved with the Rayleigh method. Numerical results from the analyses reveal different level of sensitivity of the vibration response of the SLGS to changes in the control parameters of the model. Pareto charts of the effect of the control parameters in the mathematical model show the shear modulus of the foundation to have the most dominant influence on the fundamental natural frequency of the SLGS. The small-scale effect is found to have more pronounced effect on the vibration response of the SLGS when it is fully clamped at the four edges.

---

[*] Corresponding author. Fax: +65 67911859, E-mail: kham0002@e.ntu.edu.sg (K.B. Mustapha)

**1. Introduction**

The relevance and importance of graphene is aptly summed up in the words of Geim and Novoselov [1] that "Graphene is a rapidly rising star on the horizon of materials science and condensed-matter physics. This strictly two-dimensional material exhibits exceptionally high crystal and electronic quality, and, despite its short history, has already revealed a cornucopia of new physics and potential applications." Although, it is said to be the thinnest electronic material to date [2], the exceptional properties of graphene rivals that of carbon nanotubes (CNTs). In terms of structure, graphene is described as a one-atom-thick two-dimensional (2D) layer of $Sp^2$-bonded carbon arranged in a honeycomb lattice [1, 3]. In line with the explosion of research endeavors in the fields of nanotechnology, molecular-scale engineering and quantum technology [4], graphene sheets (GSs) have also become the focus of studies in computational multiscale modeling, theoretical and experimental nano-mechanics over the recent years.

A detailed presentation of the production techniques, properties and potential applications of GSs are exhaustively documented in the recent works of Soldano et al. [5]. The high carrier mobility and the ballistic transport phenomenon of graphene have already been employed in the design of gigahertz transistors [2]. Recently, a method for the fabrication of graphene-polymer composites through the complete exfoliation of graphite and molecular-level dispersion of chemically modified GS in the polymer matrix was presented by Stankovich et al. [6]. Applications in solar cells [7], surface acoustic wave gas sensors [8], and supercapacitors for energy storage [9] have all been reported. The exceptional antibacterial property of GSs has also been revealed in the investigation by Hu et al. [10]

Meanwhile, a review of the literature on the exploration of the intrinsic properties of GSs shows that a large volume of work has been done, mostly by using the density functional theory and the tight-binding molecular dynamics, on the electrical characterization of GSs by different researchers [11-13]. Results from these studies have identified the anti-ferromagnetic ground state of some GSs; similarly, the electrical property of GS is reported to be determined by quantum confinement and edge effect. On the other hand the mechanical properties like vibration response and stability analyses of GSs remain largely inchoate [14]. Since GS is the basic building block for many of the carbon based nanostructures like CNTs, nanorings and fullerenes [15], a robust understanding of their electrical and mechanical behaviors is critical to their deployment in future applications.

Due to the immense computational requirement of molecular-level simulations, early studies on mathematical modeling of GSs (for mechanical property examination) have adopted the structural mechanics-based classical continuum theory [16-18]. While these studies have provided some insight into the mechanical behavior of GSs, the constitutive assumption in the classical continuum models that the stress at a point is only a function of the strain at the same point has raised legitimate doubt about its application beyond the microscopic realm [19]. In particular, the applicability of the classical continuum theory to the modeling of nano-level structures has revealed: (i) its under-prediction of the deflection of the nanostructures; and (ii) its over-prediction of the natural frequencies and buckling stresses of nanolevel structures [20-22]. To overcome the shortcomings of the classical continuum theory, the theory of nonlocal elasticity was advanced by Eringen [23-24] and Eringen and Edelen [25]. The studies by Eringen renew the suitability of the theory of continuum elasticity for modeling of nano-level structures. In the nonlocal continuum theory basically, the information about the long-range forces between atoms and the internal length scale is accounted for by considering the stress at a reference material point as a function of the strain states of all points in the body [26-27]. Years after the nonlocal elasticity theory is proposed, the possibility of its application it in the field of nanotechnology was reported by Peddieson et al. [28]. Series of works have since employed the theory for static and dynamic analyses of nanostructures. Among the available studies, a selective list of recent papers that provide further references on the subject include the works of Shakaee-Pour et al. [29] and Hashemnia et al. [30] on the vibration of GSs based on the molecular structural mechanics. Shakaee-Pour et al. [29] discerned that the chirality and the aspect ratio of the GS do not have a pronounced impact on its fundamental frequencies at lower modes of vibration except at the higher modes. The wave propagation characteristics of a monolayer GS based on the nonlocal elasticity theory is studied in the works of Narendar et al. [31] and Wang et al. [32]. The authors found that the nonlocal parameter introduces certain band gap region in the in-plane and flexural wave modes. Pradhan and Phadikar [33] and Pradhan and Kumar [34] studied the vibration behavior of multi-layered GSs (MLGSs) using the nonlocal plate model. The authors concluded that the nonlocal effect decreases with increase in the stiffness of the embedded MLGS. Murmu and Pradhan [35] presented the frequency and stability analyses of a simply-supported monolayer GS using the differential quadrature method (DQM).

Vibration analysis of GSs and CNTs is imperative in the design and modeling of a number of nanomechanical systems such as nano-oscillators, mass detectors, and sensing devices. In addition, vibration is an inherent part of ultrasonication in composites processing involving GSs as a reinforcing material [30]. Furthermore, it is of practical importance and interest to understand the vibration response of a single-layer GS (SLGS) combined with the effect of the surrounding medium, a situation that occurs when GSs are used in composites. The motivation for this work is predicated on the observation that the coupling effects of the residual stress state and the small-scale effect on the vibration characteristics of GSs embedded on a nonlocal elastic medium have not been reported. The investigation of the effects of the uniaxial residual stress and the stiffness of the surrounding matrix is very crucial to the design of efficient load-bearing composites.

Motivated by the above observation, the study of the transverse vibration of a pre-stressed SLGS embedded in an elastic matrix is carried out. The SLGS is modeled as a nonlocal Kirchhoff plate (NKP) which takes into account the thinness and the small-scale effect of the SLGS. The elastic matrix is structurally represented as a Pasternak foundation with two parameters. In the literature, the influence of a surrounding elastic matrix is usually captured with either the Winkler-type or the Pasternak-type elastic foundation. However, preference is given to the Pasternak-type foundation in this study based on the argument of Kerr [36] that the behavior of a large class of foundation materials occurring in practice cannot be described by the Winkler model. This is due in particular to the inconsistency in the Winkler model arising from the discontinuity of the displacements on the boundary of a uniformly loaded surface area.

## 2. Theoretical development

*2.1 The nonlocal continuum theory of the SLGS*

The formulation of the mathematical modeling of the SLGS starts off with the consideration of the continuum equivalent of the discrete configuration of the SLGS shown in Figure 1(a). Figures 1(b) and 1(c) show the schematic of the SLGS with the deformable foundation.

**Figure 1** The configuration of a SLGS being considered, (a) discrete model of the SLGS (313 carbon atoms) along with its Brillouin zone, (b) the schematic of a SLGS embedded in an elastic matrix, (c) the side view of the embedded SLGS.

In the continuum equivalent, the small-scale effects, which has been reported to have significant influence on the behavior of nano-scale materials, is included by using the theory of nonlocal continuum mechanics [23, 25]. The fundamental premise of the nonlocal theory is the assumption that the stress at a reference material point is a function of the strain at every point in the continuum formed by the material. Hence, the mathematical model of the SLGS is derived by considering a rectangular SLGS with a width ($l_1$) and a length ($l_2$) as shown in Figure 2. If we consider an element of the SLGS of mass $dm$ occupying a volume $dv$ at an arbitrary point in the domain $\forall$ of the SLGS, then depending on a displacement field $u_k$ and a velocity field $V_k$ of the element and the local acceleration of transport, the integral form of the global Newton's third law for the balance of force and moment on the element is written as [37]:

$$\oiint T_k \, dA + \iiint f_k \, dv = \iiint \rho \frac{DV_k}{Dt} \, dv \qquad (1)$$

where $T_k$ is the surface traction on the element, $f_k$ is the body force per unit volume and $\rho$ is the mass density. The term $DV_k/Dt$ is the material derivative and it approaches $\ddot{u}_k$ for a very small deformation of the element. Applying the well-known Cauchy's formula and the Gauss' theorem, the above equation reduces to:

$$\oiiint (\tau_{ik,k} + f_k - \rho \ddot{u}_k) dv = 0 \tag{2a}$$

**Figure 2** The schematic of the loading, geometry and the boundary condition of the SLGS (a) configuration of the pre-stressed SLGS, (b) geometry of the SLGS, (c) SLGS with SSSS boundary condition, and (d) SLGS with CCCC boundary condition.

In the case of an arbitrary domain, it is easily concluded from Eq. (2a) that at any material point in the continuum being considered, the following relation is true:

$$\tau_{ik,k} + f_k - \rho \ddot{u}_k = 0 \tag{2b}$$

Meanwhile, in the nonlocal continuum theory, $\tau_{ik}$ is no longer the conventional Eulerian stress tensor but the nonlocal stress tensor defined by Eringen [24] as:

$$\tau_{ik}(\boldsymbol{x}) = \int \theta(|\boldsymbol{x},\boldsymbol{x}'|,\alpha) C_{ikmn} \varepsilon_{mn}(\boldsymbol{x}') dv(\boldsymbol{x}') \quad \forall x \in v \tag{3}$$

where $C_{ikmn}$ and $\varepsilon_{mn}$ are the fourth order elasticity tensor and the strain tensor respectively. A critical part of Eq. (3) is the expression $\theta(|\boldsymbol{x},\boldsymbol{x}'|,\alpha)$ called the nonlocal kernel. The property of this kernel is exhaustively discussed by Eringen [24]. On the other hand, $|\boldsymbol{x},\boldsymbol{x}'|$ represents a distance in the Euclidean space, while $\alpha$ is a pertinent material constant defined as $e_0 a/\ell$. This material constant depends, among other things, on the lattice structure, granular size of the material, the distance between the C-C bonds (for SLGSs and CNTs) and the external characteristics length. To circumvent the problem of dealing with the kernel function in its integral form, an approximate, admissible, non-integral kernel function is required. This admissible non-integral kernel function is then used to convert the integro-partial differential equations contained in Eqs. (2) – (3) into the corresponding partial differential equations. A simplified differential operator that has been adopted to date for the kernel function in most nonlocal constitutive formulation as evidenced in [38-39] is given by $\mathcal{L} = 1 - (e_0 a)^2 \nabla^2$. In the case of a 2-dimensional structure like the SLGS, the nonlocal constitutive relations therefore become:

$$\tau_{xx} - (e_0 a)^2 \nabla^2 \tau_{xx} = \frac{E}{(1-\vartheta^2)}(\varepsilon_{xx} + \vartheta \varepsilon_{yy}) \tag{4a}$$

$$\tau_{yy} - (e_0 a)^2 \nabla^2 \tau_{yy} = \frac{E}{(1-\vartheta^2)}(\varepsilon_{yy} + \vartheta \varepsilon_{xx}) \tag{4b}$$

$$\tau_{xy} - (e_0 a)^2 \nabla^2 \tau_{xy} = G\gamma_{xy} \tag{4c}$$

in which $\tau_{xx}$ and $\tau_{yy}$ are the normal stresses in the sides parallel to the $y$ direction and the $x$ direction respectively. $\tau_{xy}$ is the shear stresses on the surfaces of the 2D continuum. $E$, $\vartheta$ and $G$ are the material's Young modulus, Poisson's ratio and the modulus of rigidity. Furthermore, the nonlocal resultant bending moments ($M_{ik}$) and the nonlocal resultant forces ($N_{ik}$) on the SLGS are related to the nonlocal parameters as:

$$M_{ik} - (e_0 a)^2 \nabla^2 M_{ik} = M_{ik}{}^L \tag{5a}$$

$$N_{ik} - (e_0 a)^2 \nabla^2 N_{ik} = N_{ik}{}^L \tag{5b}$$

where $i, k = 1, 2$. $M_{ik}{}^L$ and $N_{ik}{}^L$ are the local resultant bending moments and the local resultant forces respectively that are clearly defined in a number of literatures (e.g. [40]). Meanwhile, the typical 2D equilibrium equation of a plate undergoing a flexible transverse deformation ($w$) under a uniaxial load ($N_r$) on an elastic foundation exerting a reaction ($F_p$) is:

$$\frac{\partial q_x}{\partial x} + \frac{\partial q_y}{\partial y} + N_r \nabla^2 w = \rho h \ddot{w} + F_p \tag{6}$$

$$q_x = \frac{\partial M_{xx}}{\partial x} + \frac{\partial M_{xy}}{\partial y}, \quad q_y = \frac{\partial M_{xy}}{\partial x} + \frac{\partial M_{yy}}{\partial y}, \quad F_p = K_w w - K_g \nabla^2 w \tag{7}$$

For a SLGS in the form of a nanoplate, the nonlocal constitutive relations in Eq. (5) is employed bearing in mind the definition of $M_{ik}{}^L$, to arrive at the following elastodynamics governing equation of the pre-stressed SLGS on a nonlocal elastic medium as:

$$D\nabla^2\nabla^2 w + K_w w - K_g \nabla^2 w - (e_0 a)^2 \nabla^2 [\rho h \ddot{w} + K_w w - K_g \nabla^2 w + N_r \nabla^2 w] = \rho h \ddot{w} + N_r \nabla^2 w \tag{8}$$

Eq. (8) is now the fourth-order partial differential equation of the elliptic-type with constant coefficients. The equation governs the dynamics behavior of a uniaxially pre-stressed NKP embedded in a nonlocal deformable medium of the Reissner-type. The operator $\nabla^2$ in Eq. (8) is the Laplace operator defined as $\left(\frac{\partial}{\partial x^2} + \frac{\partial}{\partial y^2}\right)$. $\ddot{w}$ is the second time derivative of the mid-plane deflection of the SLGS, while $K_w$, $K_g$ and $h$ are the Winkler foundation modulus, the shear layer between the thickness of the SLGS and the stiffness of the surrounding elastic medium and the thickness of the nanoplate. $N_r$ is the intensity of the axially applied force and it is associated with $N_x$ and $N_y$ depicted in Figure 2(a). The solution of the nonlocal elliptic equation governing the transverse vibration response of the SLGS is sought through eigenvalue analysis. Finding the solution of the elliptic governing equation for the now well-studied local plate model, classical Kirchhoff plate (CKP), is itself not trivial and hundreds of studies have been devoted to such a case. Since the nonlocal elliptic equation governing the behavior of the SLGS does not lend itself to exact method easily, the energy-based Rayleigh method [41-42] is employed to evaluate the transverse vibration response of the SLGS through eigenvalue analysis.

### 3. The Rayleigh method

The Rayleigh method requires the formation of the functional $\Pi$ from the governing equation of the SLGS (or directly from its energy expressions), where the functional $\Pi$ is defined as:

$$\Pi = U_{max} - T_{max} \tag{9}$$

The terms $U_{max}$ and $T_{max}$ are the maximum potential and kinetic energy terms of the nonlocal nanoplate model. As it is often a good practice to cast the governing equation into the non-dimensional form, we introduce the following non-dimensional parameters:

$$\xi = x/l_1, \eta = y/l_2, \phi = l_1/l_2, \mu = e_0 a/l_1, \bar{N}_r = N_r l_1^2/D, \bar{K}_g = K_g l_1^2/D, \bar{K}_w = K_w l_1^4/D$$

$$\tag{10}$$

With the non-dimensional parameters, the explicit expressions for the maximum energy terms become:

$$U_{max} = \frac{D}{2}\int_0^1\int_0^1 \left\{\left(\frac{\partial^2 w}{\partial \xi^2} + \frac{\partial^2 w}{\partial \eta^2}\right)^2 + \bar{K}_w w^2 - (\bar{K}_g + \bar{N}_r)\left[\left(\frac{\partial w}{\partial \xi}\right)^2 + \left(\frac{\partial y}{\partial \eta}\right)^2\right] - \mu^2\left[\bar{K}_w\left[\left(\frac{\partial w}{\partial \xi}\right)^2 + \left(\frac{\partial y}{\partial \eta}\right)^2\right] + (\bar{K}_g + \bar{N}_r)\left(\frac{\partial^2 w}{\partial \xi^2} + \frac{\partial^2 w}{\partial \eta^2}\right)^2\right]d\xi d\eta\right\} \quad (11)$$

$$T_{max} = \frac{\omega^2}{2}\int_0^1\int_0^1\left\{\rho h w^2 - \mu^2 \rho h\left[\left(\frac{\partial w}{\partial \xi}\right)^2 + \left(\frac{\partial y}{\partial \eta}\right)^2\right]d\xi d\eta\right\} \quad (12)$$

Furthermore, the Rayleigh solution procedure requires that the field variable, in this case the mid-plane deflection of the SLGS be expressed in terms of an assumed mode shape function in the form of a series as:

$$w(\xi,\eta) = \sum_{m=1}^{M}\sum_{n=1}^{N} p_{mn}\varphi_m(\xi)\beta_n(\eta) \quad (13)$$

such that $p_{mn}$ is the amplitude of the function, while $\varphi_m(\xi)$ and $\beta_n(\eta)$ are the mode functions that satisfy the *Dirichlet* boundary conditions of the nanoplate. The following comparison functions are used for the assumed mode shape for the two boundary conditions considered in this study:

SSSS:

$$w(\xi,\eta) = p_{mn}\sin(m\pi\xi)\sin(n\pi\eta) \quad (14)$$

CCCC:

$$w(\xi,\eta) = p_{mn}\sin(m\pi\xi)\sin(\pi\xi)\sin(\pi\eta)\sin(n\pi\eta) \quad (15)$$

where $m = 1,2,3\ldots$; and $n = 1,2,3\ldots$;

By using Eqs. (14) – (15) in Eq. (9) bearing in mind Eqs. (11) and (12), one arrives at eigenproblem that is far more complex than the one encountered in the one-dimensional systems (e.g. CNTs and nanorods) emerges. In the next section, detail numerical results are provided to evaluate the influence of different factors on the dynamic response of the embedded SLGS.

## 4. Numerical results and discussion

Based on the formulation presented in the preceding sections, the vibration properties of the SLGS are now discussed in this section. The effective material properties of the SLGS used in the numerical evaluation of SLGS is the same as those used in Wang et al. [43] with a Young's modulus ($E$) of $1.06\ TPa$, mass density ($\rho$) of $2250\ kg/m^3$. The Poisson's ratio ($\vartheta$) of 0.25 and a thickness value ($h$) of $0.34\ nm$ are used in addition to the effective material properties.

To investigate the effect of the scale parameter, pre-stress state and the elastic matrix properties on the vibrations of SLGS, the results including and excluding the nonlocal parameter, pre-stress and the foundation parameter are compared. The ratio of the computational results with the nonlocal parameter to those based on the classical (local) theory is given by

$$R_{nm} = (\lambda_{nm})_{NL} / (\lambda_{nm})_L \tag{16}$$

where $(\lambda_{nm})_{NL}$ is the natural frequency of the SLGS from the nonlocal theory, while $(\lambda_{nm})_L$ is the corresponding natural frequency from the classical theory. This ratio provides us with the information about the degree of influence of the nonlocal parameter. $\lambda_{nm}$ is the non-dimensional natural frequency parameter defined as $\lambda_{nm} = \omega_{nm} l_1^2 \sqrt{\rho h/D}$ and $D = E/12(1-\vartheta^2)$. Likewise, the ratio of the natural frequencies of the embedded SLGS with that of the unembedded SLGS is given by:

$$\chi = \frac{\text{Natural frequency of the embedded SLGS } (\lambda_{nm})_{embedded}}{\text{Natural frequency of the unembedded SLGS } (\lambda_{nm})_{unembedded}} \tag{17}$$

*4.1 The effect of the half-wave numbers ($n, m$) on the vibration mode of the SLGS*

The natural frequency of the SLGS and its mode shapes during the free vibration analysis depend on a number of factors. Of these factors, the duo of the half-wave numbers ($n, m$) and the boundary condition generally determines the mode to which the SLGS will be subjected. In Figures 3 and 4, we present the modes of vibration of a SLGS for the first six natural frequencies in the case of SSSS and CCCC boundary conditions at a small-scale coefficient $\mu = 0.4\ nm$.

**Figure 3** The vibration modes of a SLGS based on a nonlocal Kirchhoff nanoplate under the SSSS boundary condition.

**Figure 4** The vibration modes of a SLGS based on a nonlocal Kirchhoff nanoplate under a fully clamped (CCCC) boundary condition.

The difference in the vibration modes of the SLGS under the two boundary configurations is quite obvious from Figures 3 and 4. It should be noted that, the natural frequency values predicted by the classical continuum theory of plate are higher than those indicated in Figures 3 and 4. The small-scale effect taken into the consideration in the present analysis allows the consistent inclusion of the size effect in the mathematical model of the nanoplate. The edges of the SLGS in Figure 3 are observed to bulge with some very small amplitude as a result of the rotational effect that is left unconstrained in the case of the SSSS boundary configuration. However, the edges are fully clamped in Figure 4 as observed in the first set of rectangular mesh at each of the four edges. The sides of these rectangular meshes in Figure 4 are seen maintaining a parallel position to the undeflected surface of the SLGS. The CCCC boundary condition is a suitable boundary condition in nanocircuits in which GSs are employed as the base.

*4.2 The effect of the nonlocal parameter*

It has to be stated that the magnitude of the nonlocal parameter $\mu$ determines the influence of the small-scale effect in all the analyses. In the definition of $\mu$ (Eq. (10)), there are two constants that make up its definition: $e_0$ and $a$. According to Eringen [24], $e_0$ is a constant appropriate to specific materials. The value of $e_0$ is determined through an experimental phonon dispersion curve [44]. The determination of the exact value of $e_0$ remains a challenge for the interdisciplinary researchers working on SLGSs and CNTs. For example, Zhang et al. [45] estimated the value of $e_0$ for a SWCNT by curve fitting the theoretical results obtained using nonlocal continuum theory to those gotten from MD simulations for the axial buckling strain of the SWCNT. On the other hand, Zhang et al. [46] presented analysis of elastic interaction between Stone–Wales and divacancy defects on a GS. More experimental tests are still required to determine $e_0 a$ more accurately for SLGSs. However, there seemed to be a consensus of opinions on the fact that a conservative value of the nonlocal parameter should be in the range of $0 - 2\ nm$ [35, 47].

The relationship between the frequency ratios $(R_{nm})$ and the small-scale parameter for five combinations of the half-wave mode numbers $(n, m)$ for a SLGS of dimensions ($10\ nm\ \times 10\ nm\ \times 0.34\ nm$) under the SSSS and the CCCC boundary conditions is shown in Figure 5. The case of a SSSS SLGS is shown in Figure 5(a), while Figure 5(b) reveals the case of the CCCC SLGS. The frequency ratio $(R_{nm})$ is an ideal quantitative index to assess the nature of the influence of the small-scale effect on the free vibration response of a SLGS. It is clearly seen from Figures 5(a) and 5(b) that for both cases of SSSS and CCCC, the frequency ratios $(R_{nm})$ are less than unity. This implies that the natural frequencies of the SLGS obtained based on the classical continuum model are usually over-predicted if the small length scale effect between the individual carbon atoms in GS is neglected. In addition, the reduction in the values of the natural frequency is even more pronounced for higher vibration modes of the SLGS.

Due to the similarity in Figures 5(a) and 5(b) for the two boundary configurations of the SLGS, what is not directly observable in these figures is revealed in Tables 1 and 2. Tables 1 and 2 contain both the non-dimensional natural frequency values and the frequency ratios $(R_{nm})$ for the SSSS and the CCCC SLGSs. The SLGS is not embedded and not under the influence of any uniaxial stress. Table 1 serves two purposes: (i) to serve as a validation of our solution procedure, and (ii) to reveal the actual value of the natural frequency parameter for different values of the nonlocal parameter in case of the a SLGS under the SSSS boundary condition. By looking at the values of the first seven $R_{nm}$ in Table 1 and comparing with the values obtained by Lu et al. [39], where the $R_{nm}$ values for a nanoplate with a SSSS boundary condition is presented, a very good agreement is observed. In addition, by comparing the values of the natural frequency of the SLGS in Table 1 (SSSS) and in Table 2 (CCCC), it becomes very glaring that the nonlocal parameter (small-scale coefficient) has more effect on the frequency of the SLGS when it is fully clamped than when it is simply supported on all edges. The same behavior was observed by these authors in their work on embedded CNTs [48].

**Figure 5** Variations of the frequency ratios ($R_{nm}$) with the small-scale coefficient ($\mu$) for different values of the half-wave mode number, (a) SSSS boundary condition, (b) CCCC boundary condition.

**Table 1** Non-dimensional frequencies and frequency ratios of the SLGS with SSSS boundary condition.

**Table 2** Non-dimensional frequencies and frequency ratios of the SLGS with CCCC boundary condition.

*4.3 The influence of the elastic medium on the behavior of the SLGS*

The changes in the natural frequencies of the SLGS when it is in contact with the surrounding elastic medium reflect the changes in its dynamic behavior under the influence of the surrounding elastic medium. In the mathematical model presented in section 2, the incorporation of the elastic medium into the mathematical model of the unembedded SLGS comes through the use of the Pasternak model. The Pasternak model is characterized by two parameters: the Winkler modulus and the shear modulus. The Winkler modulus represents the stiffness of the assumed elastic matrix in which the SLGS is embedded. We first present the influence of the stiffness (Winkler modulus) of an elastic matrix on the frequency response of the SLGS.

Figures 6(a) and 6(b) show the influence of the surrounding medium stiffness on the natural frequency parameter of the SLGS for the two support configurations of SSSS and CCCC. The graphs in Figures 6(a) and 6(b) are presented in terms of the normalized natural frequency parameter ($\chi$) of the SLGS. The essence of the normalized natural frequency parameter is to provide a unified comparison for each value of the natural frequency of the embedded SLGS at different values of the small-scale parameter. The trend for the two boundary conditions shows that, irrespective of the way the SLGS is supported at each of its edges, its natural frequency will increase in an elastic medium. The level of the increase will however depend on the stiffness of the elastic medium in consideration. Furthermore, it will be observed that, the smaller the small-scale coefficient of the SLGS, the more pronounced the effect of the medium's stiffness on its natural frequency. This behavior is consistent with the fact that a very a high value of the small-scale parameter indicates a lower stiffness of the SLGS which leads to the SLGS being prone to a greater additional force from the surrounding medium.

**Figure 6** Variation of the normalized frequency ($\chi$) of the embedded SLGS with the Winkler modulus parameter $\bar{K}_w$ for different values of the nonlocal parameter, (a) SLGS with SSSS boundary condition, (b) SLGS with CCCC boundary condition.

Tables 3 and 4 provide further details in addition to Figure 6. In these tables, the actual numerical value of the natural frequency for a range of the nonlocal parameter and the Winkler modulus are presented. Tables 3 and 4 reveal the increase in the frequency values at each level of the half-wave mode numbers $(n, m)$ for different values of the stiffness of the elastic medium ($\bar{K}_w$). The numerical values for the case of a nanoplate embedded in a foundation of the Winkler-type are in general not available in the literature. However, a pointer to the accuracy of the values of the predicted natural frequencies in Tables 3 and 4 is found in the second column of Tables 3 and 4. The natural frequency in this column represents that of a plate with a zero small-scale parameter (or simply the classical Kirchhoff plate) embedded in a foundation with stiffness of $10, 100$ and $1000$. By comparing the value of the fundamental frequency ($\lambda_{11}$) of the SSSS plate (in Table 3 only) characterized with the superscript (a) predicted from this analysis with that of Matsunaga [49], a very good agreement is found. The tables further reveal that the Winkler modulus has more effect on the lower vibration modes of the SLGS and a reduced effect at the higher modes of the vibration. For example, if we consider the value of the fundamental frequency ($\lambda_{11}$) and say $\lambda_{33}$ of a simply-supported SLGS under a foundation with a non-dimensional Winkler parameter $\bar{K}_w = 10$ and $\bar{K}_w = 100$, one would observe that the rate of increase of the fundamental frequency ($\lambda_{11}$) is higher than that of the higher mode $\lambda_{33}$.

We next consider the unified influence of the Winkler modulus and the Pasternak shear modulus on the behavior of the SLGS. For this analysis, a constant value of $\bar{K}_w = 100$ is used for the non-dimensional Winkler modulus, while the nonlocal parameter is maintained at a value of $\mu = 0.2$. The other geometrical properties remain as described in the first paragraph of this section. Figure 7 shows the relationship obtained between the non-dimensional natural frequencies of the SLGS with the Pasternak shear modulus for different values of the half-wave mode number. A linear relationship exists between the vibration response of the SLGS and the shear modulus at lower and higher modes of vibration. However, the frequency-Pasternak shear modulus relationship plotted in Figure 7(a) for a SLGS under a SSSS boundary condition are lower than those obtained in the case of a fully clamped SLGS. Additionally, in the case of a fully clamped SLGS, it is observed that as the vibration mode increases, the gap between the natural frequency values decreases in the presence of the nonlocal parameter.

**Table 3** Non-dimensional frequencies of the SSSS SLGS on an elastic foundation of the Winkler-type.

**Table 4** Non-dimensional frequencies of a CCCC SLGS on an elastic foundation of the Winkler-type.

**Figure 7** The natural frequency-Pasternak shear modulus relationship for different vibration modes of the SLGS, (a) SLGS with SSSS boundary condition, (b) SLGS with CCCC boundary condition.

*4.4 The influence of pre-stress condition on the dynamic behavior of the SLGS*

There are a number of practical situations that require taking into account the effect of pre-stress in nanostructures. In particular, the effect of the uniaxial stress becomes important when SLGSs are used as reinforcement. Depending on the nature of the uniaxial stress coming from the uniaxial force, the SLGS reinforced composites will respond differently. For instance, during service use of such composites could be subjected to a wide band of temperature changes. As a result of this wide range of temperature variation, the slab of the composites might expand and compressive forces will be induced on the SLGS.

In situations where the compressive force is excessive, the SLGS reinforced composites may experience a blowup [50]. To be able to check the sensitivity of the frequency of the SLGS to changes in the values of the axial stress, the values of the parameters that are kept constant under the variation of the value of the initial stress include the Winkler modulus parameter $\bar{K}_w = 200$, and $\bar{K}_g = 10$. Additionally, the value of the small-scale parameter is varied within the range $0 \leq \mu \leq 1\ nm$.

Figures 8 and 9 present the relationship between the normalized natural frequency ($\Lambda$) and the compressive uniaxial stress for the two boundary configurations studied in this work. The normalization of the frequency in this subsection is carried out with respect to the natural frequency of the embedded SLGS without the pre-stress effect. This enables a unified quantification of the effect of the initial compressive stress on the dynamics response of the SLGS. In Figures 8 and 9, the effect of the compressive stress is observed to be more pronounced when the nonlocal parameter is high. However, the distinction between the two normalized frequency-compressive stress relationships for the SSSS and CCCC boundary conditions is obvious. For a SLGS with SSSS boundary condition that is characterized by a very high small-scale coefficient parameter, a quadratic effect of the stress on the natural frequency is observed in Figure 8. On the contrary, the fully clamped SLGS undergoes a linear behavior with closely-spaced normalized frequency-compressive stress curve.

**Figure 8** The effect of the compressive uniaxial stress on the frequency of an embedded SLGS with a SSSS boundary condition.

**Figure 9** The effect of the compressive uniaxial stress on the natural frequency of an embedded SLGS with a CCCC boundary condition.

*4.5 The combined influence of four variables on the dynamic behavior of the SLGS*

The graphs presented in the preceding subsections establish the relationships between a certain control parameter of the mathematical model and the natural frequency of the SLGS while the other parameters are kept constant. However, there are situations in practical cases when there could be variations in the values of more than one parameter at the same time. Therefore, in this subsection, we investigate the order of significance of all the control parameters that make up the mathematical model when there are multiple variations. The control parameters in the mathematical model include the nonlocal parameter $(\mu)$, the compressive stress parameter $(\bar{N}_r)$, the Winkler modulus parameter $(\bar{K}_w)$ and the Pasternak shear modulus $(\bar{K}_g)$.

In Figures 10 and 11, the Pareto charts of the effects of variations in the values of these parameters are presented for the SLGS. A Pareto chart is often used in process control and design of experiments (DOE). It becomes quite handy when it is required to provide a graphical elucidation and the relative importance of the factors whenever there are variations in the values of the factors that affect a process, products or response [51]. In our present study, we have taken the prediction of the natural frequency as a 'theoretical experiment', and the response, for the two Pareto charts presented is the fundamental natural frequency $(\lambda_{11})$. The DOE involving four factors at two levels $(2^k)$, where the superscript $k$ represents the number of variables, involves sixteen runs of the theoretical experiment.

The value of the fundamental natural frequency along with the corresponding values of the parameters at each level of the runs is entered into a commercial statistical package, MINITAB® for the generation of the Pareto charts. The chart for the SLGS with the SSSS boundary condition is shown in Figure 10. Figure 11 is the Pareto chart for the case of the SLGS with the CCCC boundary condition. It is observed from these charts that the factor with the most dominant effect on the fundamental natural frequency of the SLGS is the shear modulus of the elastic matrix. This is followed by the nonlocal parameter (i.e. the small-scale coefficient). The compressive stress then comes next in the order of significance. The order of significance of the factors on the dynamical response of the SLGS for both the SSSS and the CCCC boundary configurations is the same, but the critical red line in Figures 10 and 11 occurs at different values. In the case of the SSSS in Figure 10, it occurs at a value of $0.91$, while it occurs at a value of $0.74$ in Figure 11. Those factors that occur below this critical red line do not contribute in any significant way to the response in consideration. Some combined effects (e.g. $AB, AD, BD, ABD$) of the factors will be observed from both figures. For example, the term $ABD$ is in the parlance of the industrial experimentalists, called an interaction effect between the factors $A$, $B$ and $D$. Sometimes, the interaction effect can be more significant than the individual effect.

**Figure 10** The Pareto chart of the Winkler modulus, Pasternak shear modulus, the nonlocal factor and the compressive stress for a simply supported SLGS.

**Figure 11** The Pareto chart of the Winkler modulus, Pasternak shear modulus, the nonlocal factor and the compressive stress for a fully clamped SLGS.

**5. Conclusion**

Vibration analysis of an embedded SLGS has been carried out based on the NKP theory. The theory allows the inclusion of the small-scale coefficient of the SLGS in terms of the characteristics length between the carbon atoms of the SLGS. The sensitivity of the natural frequencies of the SLGS to different parameters contained in the elastodynamics governing equation of the SLGS is investigated.

The results show that for a SLGS with a very high size effect, the natural frequency predicted using the classical plate theory has a higher level of over-estimation. It is also observed that the reduction in the values of the natural frequency is more pronounced for higher vibration modes of the SLGS. In addition a direct relationship exists between the vibration response of the SLGS and the two moduli of the elastic matrix at lower and higher modes of vibration. However, the frequency-Pasternak shear modulus relationship for a SLGS under the SSSS and CCCC support conditions is linear.

Furthermore, the Pareto charts reveal that the factor with the most dominant effect on the fundamental natural frequency of the SLGS is the shear modulus of the foundation. This is followed by the small-scale coefficient and then the compressive stress level. It is therefore important to give special consideration to these factors in the optimal design of nanoelectronics in which the SLGS will be used and composites.

## List of Figure captions

# Tables

**Table 1** Non-dimensional frequencies and frequency ratios of the SLGS with SSSS boundary condition.

| $(m, n)$ | $\mu$ | | | | | | | | |
|---|---|---|---|---|---|---|---|---|---|
| | 0 | 0.1 | 0.2 | 0.3 | 0.4 | 0.5 | 0.6 | 0.7 | 0.8 |
| $(1,1)$ | 19.7392 | 17.1808 | 13.0726 | 10.0211 | 7.9791 | 6.5797 | 5.5786 | 4.8331 | 4.2590 |
| $R_{11}$ | 1. | 0.9139 | 0.7475 | 0.6001 | 0.4904 | 0.4105 | 0.3512 | 0.3061 | 0.2708 |
| $(1,2)$ | 49.3482 | 40.3803 | 28.6157 | 21.1552 | 16.5455 | 13.5127 | 11.3918 | 9.83416 | 8.64522 |
| $R_{12}$ | 1. | 0.8183 | 0.5799 | 0.4287 | 0.3352 | 0.2738 | 0.2308 | 0.1993 | 0.1752 |
| $(2,1)$ | 49.3483 | 40.3803 | 28.6157 | 21.1552 | 16.5455 | 13.5127 | 11.3918 | 9.83416 | 8.64522 |
| $R_{21}$ | 1. | 0.8183 | 0.5798 | 0.4287 | 0.3353 | 0.2738 | 0.2308 | 0.1993 | 0.1752 |
| $(2,2)$ | 78.9568 | 59.0222 | 38.7198 | 27.7321 | 21.3842 | 17.3378 | 14.5558 | 12.533 | 10.9989 |
| $R_{22}$ | 1. | 0.7475 | 0.4903 | 0.3512 | 0.2708 | 0.2196 | 0.1844 | 0.1587 | 0.1393 |
| $(2,3)$ | 128.3056 | 84.9152 | 51.8126 | 36.2214 | 27.6524 | 22.3093 | 18.6775 | 16.0545 | 14.0735 |
| $R_{23}$ | 1. | 0.6618 | 0.4038 | 0.2823 | 0.2155 | 0.1738 | 0.1456 | 0.1251 | 0.1097 |
| $(3,2)$ | 128.3051 | 84.9152 | 51.8126 | 36.2214 | 27.6524 | 22.3093 | 18.6775 | 16.0545 | 14.0735 |
| $R_{32}$ | 1. | 0.6618 | 0.4038 | 0.2823 | 0.2155 | 0.1738 | 0.1456 | 0.1251 | 0.1097 |
| $(3,3)$ | 177.6530 | 106.6161 | 62.3973 | 43.1014 | 32.7505 | 26.3622 | 22.0428 | 18.9325 | 16.5882 |
| $R_{33}$ | 1. | 0.6001 | 0.3512 | 0.2426 | 0.1843 | 0.1483 | 0.1240 | 0.1066 | 0.0934 |
| $(3,4)$ | 246.7412 | 132.5078 | 74.8398 | 51.2193 | 38.7818 | 31.1643 | 26.0338 | 22.3477 | 19.5731 |
| $R_{34}$ | 1. | 0.5370 | 0.3033 | 0.2075 | 0.1572 | 0.1263 | 0.1055 | 0.0906 | 0.0793 |
| $(4,4)$ | 315.8271 | 154.8792 | 85.5367 | 58.2231 | 43.9956 | 35.3201 | 29.4898 | 25.3063 | 22.1597 |

**Table 2** Non-dimensional frequencies and frequency ratios of the SLGS with CCCC boundary condition.

| $(m,n)$ | \multicolumn{9}{c}{$\mu$} |
|---|---|---|---|---|---|---|---|---|---|
|  | 0 | 0.1 | 0.2 | 0.3 | 0.4 | 0.5 | 0.6 | 0.7 | 0.8 |
| $(1,1)$ | 37.2206 | 33.1169 | 25.9785 | 20.2793 | 16.3051 | 13.5194 | 11.5003 | 9.9847 | 8.8112 |
| $R_{11}$ | 1. | 0.8897 | 0.6980 | 0.5448 | 0.4381 | 0.3632 | 0.3090 | 0.2683 | 0.2367 |
| $(1,2)$ | 76.2371 | 59.8038 | 40.7486 | 29.6176 | 22.9851 | 18.6965 | 15.7256 | 13.5558 | 11.9055 |
| $R_{12}$ | 1. | 0.7845 | 0.5345 | 0.3885 | 0.3015 | 0.2452 | 0.2062 | 0.1778 | 0.1562 |
| $(2,1)$ | 76.2372 | 59.8038 | 40.7486 | 29.6176 | 22.9851 | 18.6965 | 15.7256 | 13.5558 | 11.9055 |
| $R_{21}$ | 1. | 0.7844 | 0.5345 | 0.3885 | 0.3015 | 0.2452 | 0.2063 | 0.1778 | 0.1561 |
| $(2,2)$ | 113.3930 | 80.4437 | 50.9775 | 36.0703 | 27.6722 | 22.3792 | 18.7611 | 16.1397 | 14.1559 |
| $R_{22}$ | 1. | 0.70942 | 0.4496 | 0.3181 | 0.2440 | 0.1974 | 0.1655 | 0.1423 | 0.1248 |
| $(2,3)$ | 164.2631 | 104.2982 | 62.4355 | 43.4018 | 33.0602 | 26.6431 | 22.2924 | 19.1546 | 16.7872 |
| $R_{23}$ | 1. | 0.6349 | 0.3801 | 0.2642 | 0.2013 | 0.16220 | 0.1357 | 0.1166 | 0.1023 |
| $(3,2)$ | 164.2630 | 104.2985 | 62.4355 | 43.4018 | 33.0602 | 26.6431 | 22.2924 | 19.1546 | 16.7872 |
| $R_{32}$ | 1. | 0.6349 | 0.3801 | 0.2642 | 0.2013 | 0.1622 | 0.1357 | 0.1166 | 0.1022 |
| $(3,3)$ | 214.4235 | 124.3394 | 71.8921 | 49.4986 | 37.5644 | 30.2194 | 25.2592 | 21.6907 | 19.0022 |
| $R_{33}$ | 1. | 0.5798 | 0.3353 | 0.2308 | 0.1752 | 0.1409 | 0.1178 | 0.1012 | 0.0886 |
| $(3,4)$ | 284.1692 | 148.4415 | 83.2233 | 56.8534 | 43.0181 | 34.5573 | 28.8632 | 24.7737 | 21.6963 |
| $R_{34}$ | 1. | 0.5224 | 0.2929 | 0.2001 | 0.1514 | 0.1216 | 0.1015 | 0.0872 | 0.0763 |
| $(4,4)$ | 353.6572 | 169.4553 | 93.1236 | 63.3138 | 47.8218 | 38.3845 | 32.0444 | 27.4966 | 24.0765 |

**Table 3** Non-dimensional frequencies of the SSSS SLGS on an elastic foundation of the Winkler-type.

| $(m, n)$ | $\mu$ | | | | | | | | |
|---|---|---|---|---|---|---|---|---|---|
| | 0 | 0.1 | 0.2 | 0.3 | 0.4 | 0.5 | 0.6 | 0.7 | 0.8 |
| $\bar{K}_w = 10$ | | | | | | | | | |
| $(1, 1)$ | 19.9909[a] | 18.3142 | 15.0906 | 12.2615 | 10.1834 | 8.69786 | 7.62017 | 6.81978 | 6.2113 |
| $(1, 2)$ | 49.4492 | 40.5045 | 28.7899 | 21.3903 | 16.8454 | 13.8777 | 11.8226 | 10.3301 | 9.2054 |
| $(2, 2)$ | 79.0201 | 59.1069 | 38.8487 | 27.9119 | 21.6167 | 17.6238 | 14.8953 | 12.9258 | 11.4445 |
| $(3, 2)$ | 128.3440 | 84.9741 | 51.9091 | 36.3592 | 27.8327 | 22.5323 | 18.9433 | 16.363 | 14.4244 |
| $\bar{K}_w = 100$ | | | | | | | | | |
| $(1, 1)$ | 22.1277[a] | 20.6253 | 17.8249 | 15.5027 | 13.9177 | 12.8706 | 12.1683 | 11.6837 | 11.3393 |
| $(1, 2)$ | 50.3513 | 41.6001 | 30.3127 | 23.3996 | 19.3327 | 16.8105 | 15.1583 | 14.0254 | 13.2189 |
| $(2, 2)$ | 79.5876 | 59.8634 | 39.9903 | 29.4812 | 23.6068 | 20.0152 | 17.6599 | 16.0336 | 14.8653 |
| $(3, 2)$ | 128.6941 | 85.5023 | 52.7688 | 37.5765 | 29.4051 | 24.448 | 21.1861 | 18.9142 | 17.2645 |
| $\bar{K}_w = 1000$ | | | | | | | | | |
| $(1, 1)$ | 37.2778[a] | 36.4061 | 34.8959 | 33.7688 | 33.0712 | 32.6443 | 32.3739 | 32.1949 | 32.0715 |
| $(1, 2)$ | 58.6108 | 51.2891 | 42.6481 | 38.0466 | 35.6897 | 34.3888 | 33.6121 | 33.1166 | 32.7832 |
| $(2, 2)$ | 85.0542 | 66.9599 | 49.9922 | 42.0603 | 38.1744 | 36.0638 | 34.8119 | 34.0158 | 33.481 |
| $(3, 2)$ | 132.1447 | 90.6123 | 60.7004 | 48.0832 | 42.0078 | 38.7002 | 36.7267 | 35.4647 | 34.6131 |

[a]Classical plate frequency coincides with the prediction of Matsunaga [49]

**Table 4** Non-dimensional frequencies of a CCCC SLGS on an elastic foundation of the Winkler-type.

| $(m,n)$ | $\mu$ | | | | | | | | |
|---|---|---|---|---|---|---|---|---|---|
| | 0 | 0.1 | 0.2 | 0.3 | 0.4 | 0.5 | 0.6 | 0.7 | 0.8 |
| $\bar{K}_w = 10$ | | | | | | | | | |
| $(1,1)$ | 37.3547 | 33.2675 | 26.1703 | 20.5243 | 16.6089 | 13.8843 | 11.9272 | 10.4735 | 9.36149 |
| $(1,2)$ | 76.3025 | 59.8874 | 40.8711 | 29.786 | 23.2016 | 18.962 | 16.0404 | 13.9198 | 12.3183 |
| $(2,2)$ | 113.4371 | 80.5058 | 51.0755 | 36.2087 | 27.8523 | 22.6013 | 19.0257 | 16.4465 | 14.5048 |
| $(3,2)$ | 164.2935 | 104.346 | 62.5155 | 43.5169 | 33.2111 | 26.8301 | 22.5155 | 19.4139 | 17.0823 |
| $\bar{K}_w = 100$ | | | | | | | | | |
| $(1,1)$ | 38.5405 | 34.5937 | 27.8367 | 22.6108 | 19.1273 | 16.8159 | 15.2487 | 14.1313 | 13.3281 |
| $(1,2)$ | 76.8912 | 60.6341 | 41.9577 | 31.2603 | 25.0662 | 21.2028 | 18.6358 | 16.8452 | 15.548 |
| $(2,2)$ | 113.8337 | 81.0629 | 51.9491 | 37.4309 | 29.4236 | 24.51162 | 21.2598 | 18.9865 | 17.3317 |
| $(3,2)$ | 164.5672 | 104.7765 | 63.2312 | 44.5392 | 34.5395 | 28.4583 | 24.4326 | 21.6078 | 19.5398 |
| $\bar{K}_w = 1000$ | | | | | | | | | |
| $(1,1)$ | 48.8403 | 45.7934 | 40.9253 | 37.5666 | 35.5788 | 34.3915 | 33.649 | 33.1616 | 32.8274 |
| $(1,2)$ | 82.5353 | 67.6498 | 51.5796 | 43.3267 | 39.0936 | 36.7363 | 35.317 | 34.4058 | 33.7897 |
| $(2,2)$ | 117.7212 | 86.4361 | 59.9892 | 47.9695 | 42.0208 | 38.7404 | 36.7693 | 35.5034 | 34.6466 |
| $(3,2)$ | 167.2790 | 108.986 | 69.987 | 53.7003 | 45.749 | 41.3504 | 38.6904 | 36.9716 | 35.8023 |

**Figure1a**
**Click here to download high resolution image**

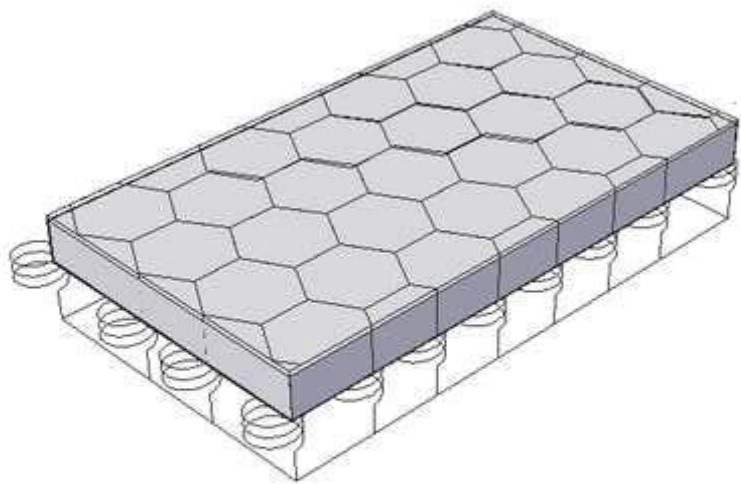

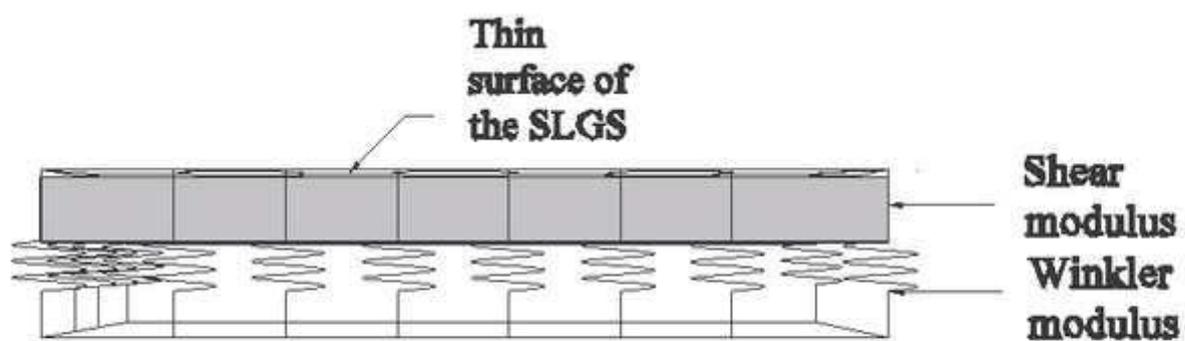

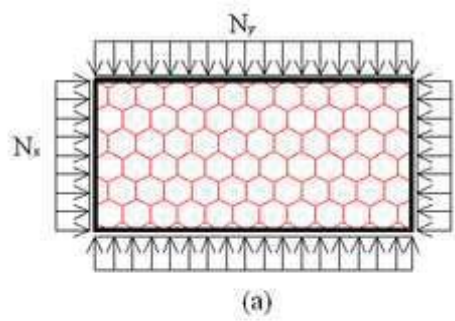
(a)

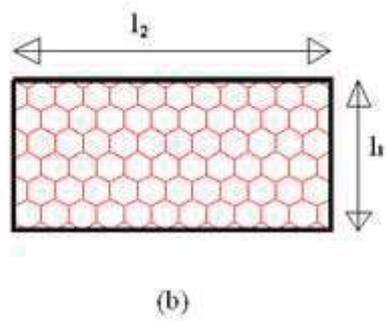
(b)

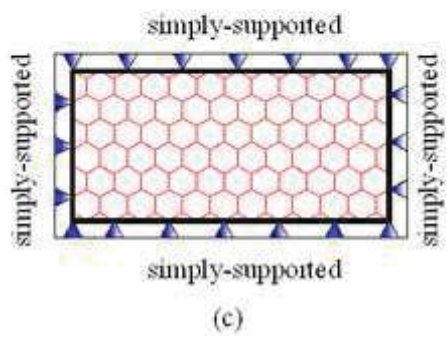
(c)

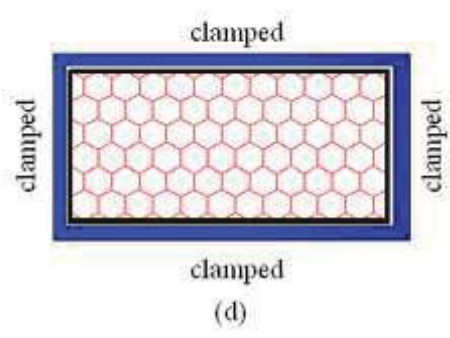
(d)

**Figures 3&4**

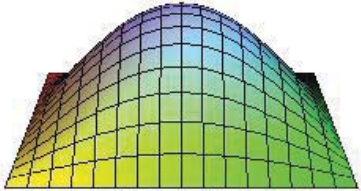 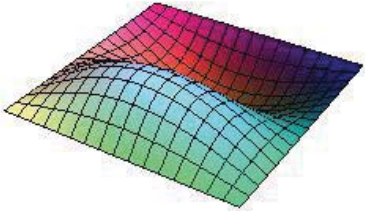 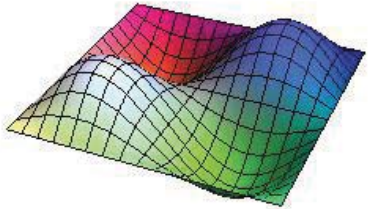

$\lambda_{11} = 7.9791$        $\lambda_{12} = 16.5455$        $\lambda_{22} = 21.3842$

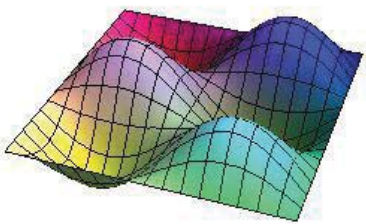 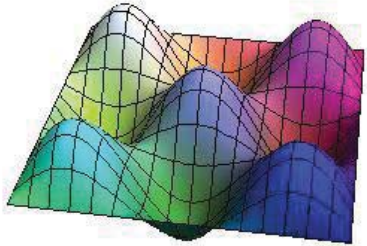 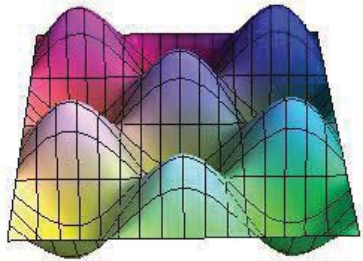

$\lambda_{23} = 27.6524$        $\lambda_{33} = 32.7505$        $\lambda_{34} = 38.7818$

**Figure 3**. The vibration modes of a SLGS based on a nonlocal Kirchhoff nanoplate under the SSSS boundary condition.

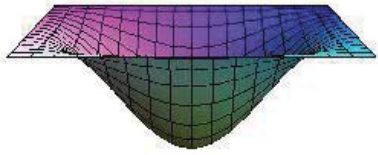 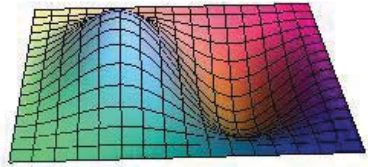 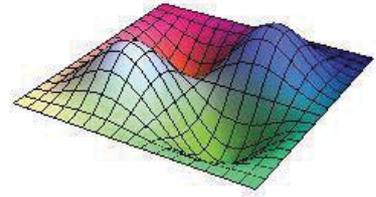

$\lambda_{11} = 16.3051$        $\lambda_{12} = 22.9851$        $\lambda_{22} = 27.6722$

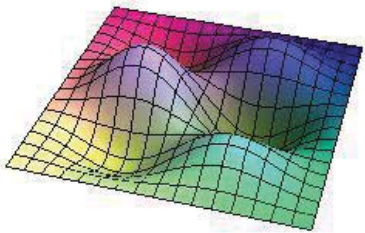 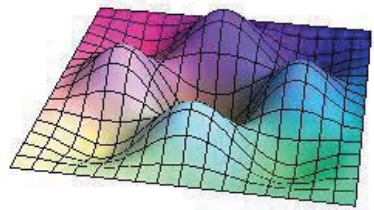 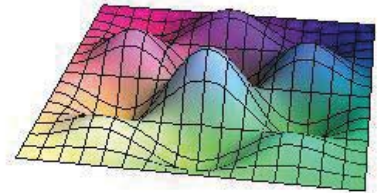

$\lambda_{23} = 33.0602$        $\lambda_{33} = 37.5644$        $\lambda_{34} = 43.0181$

**Figure 4**. The vibration modes of a SLGS based on a nonlocal Kirchhoff nanoplate under a fully clamped (CCCC) boundary condition.

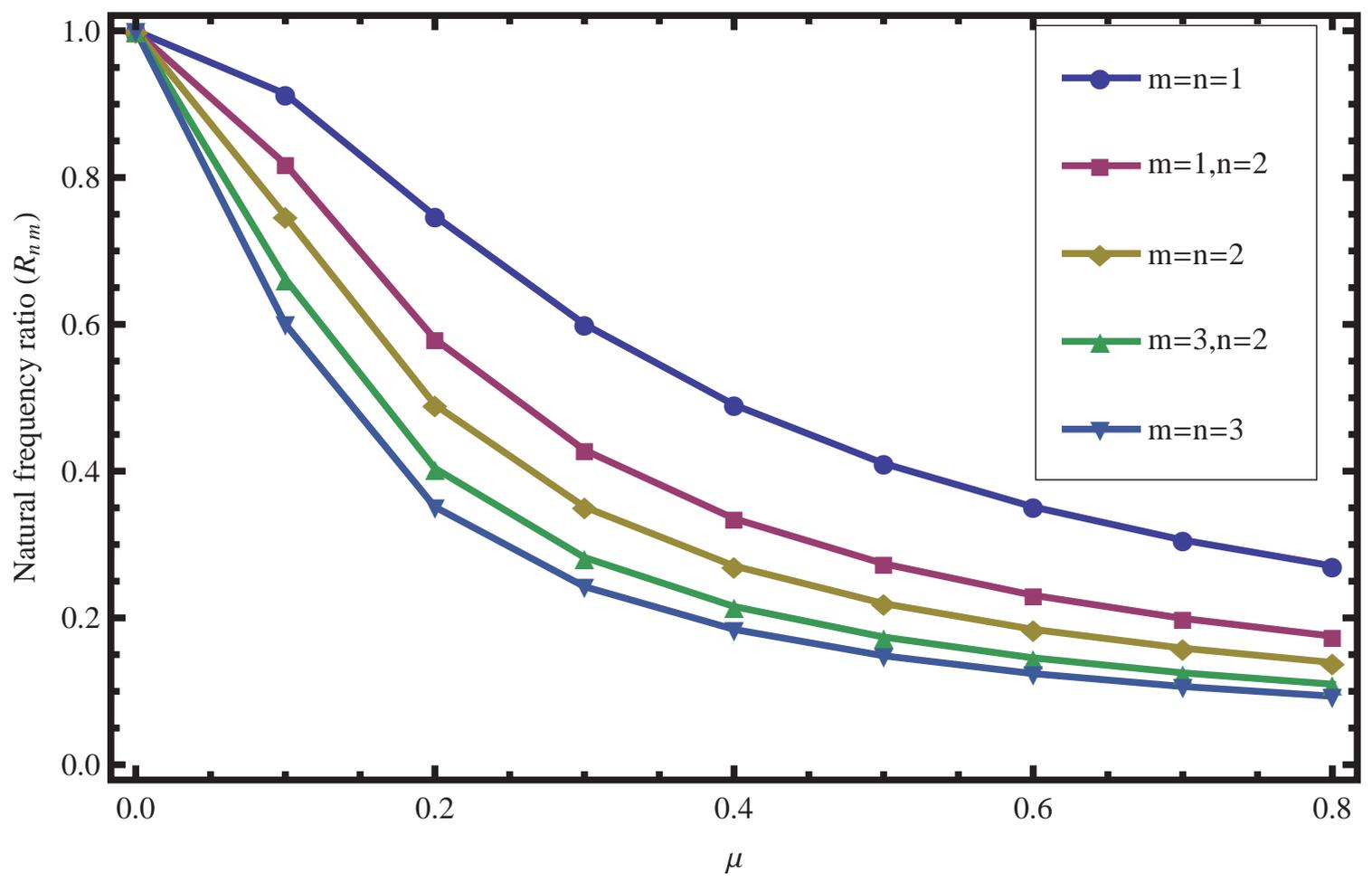

**Figure 5b**

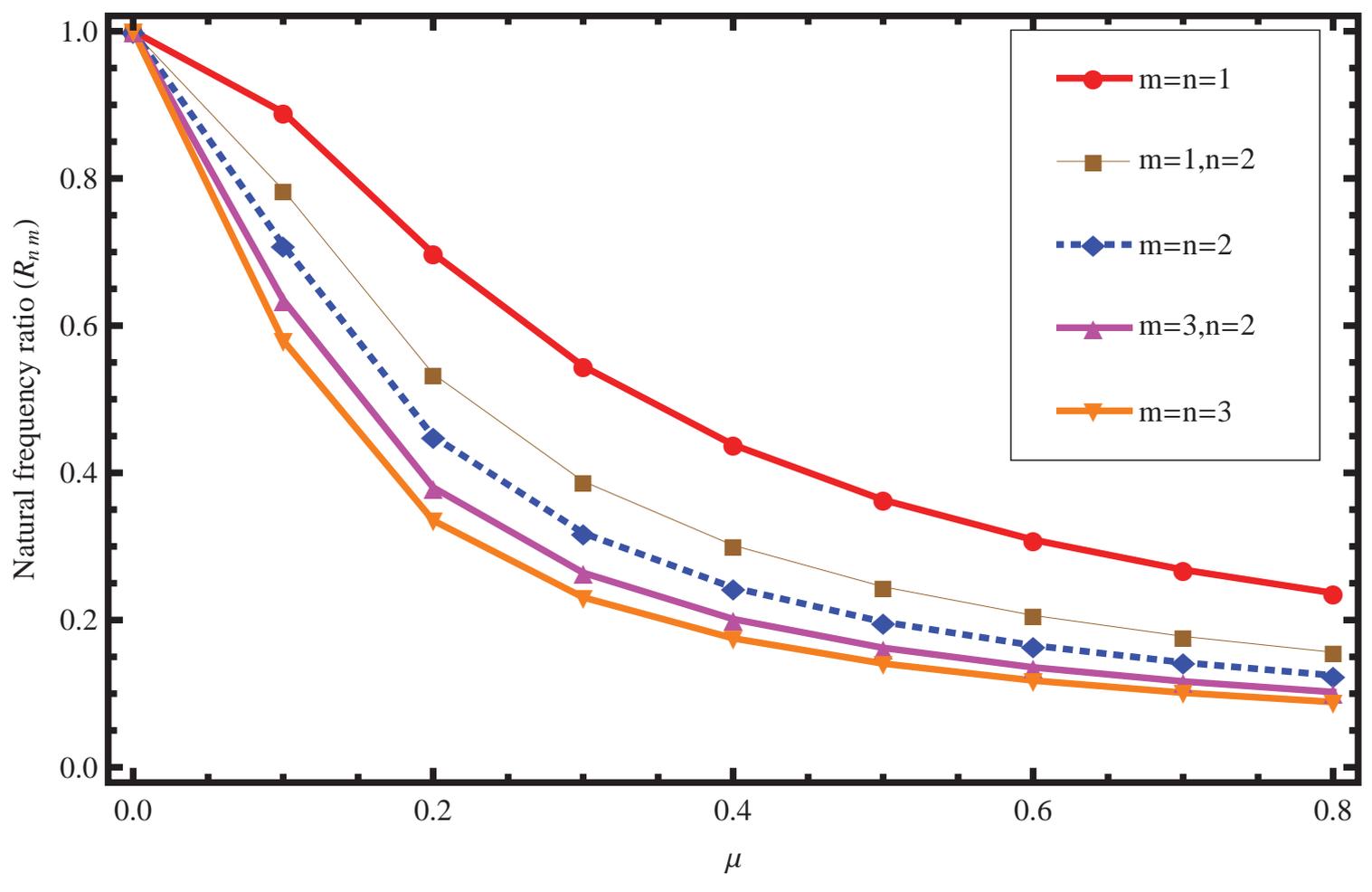

**Figure 6a**

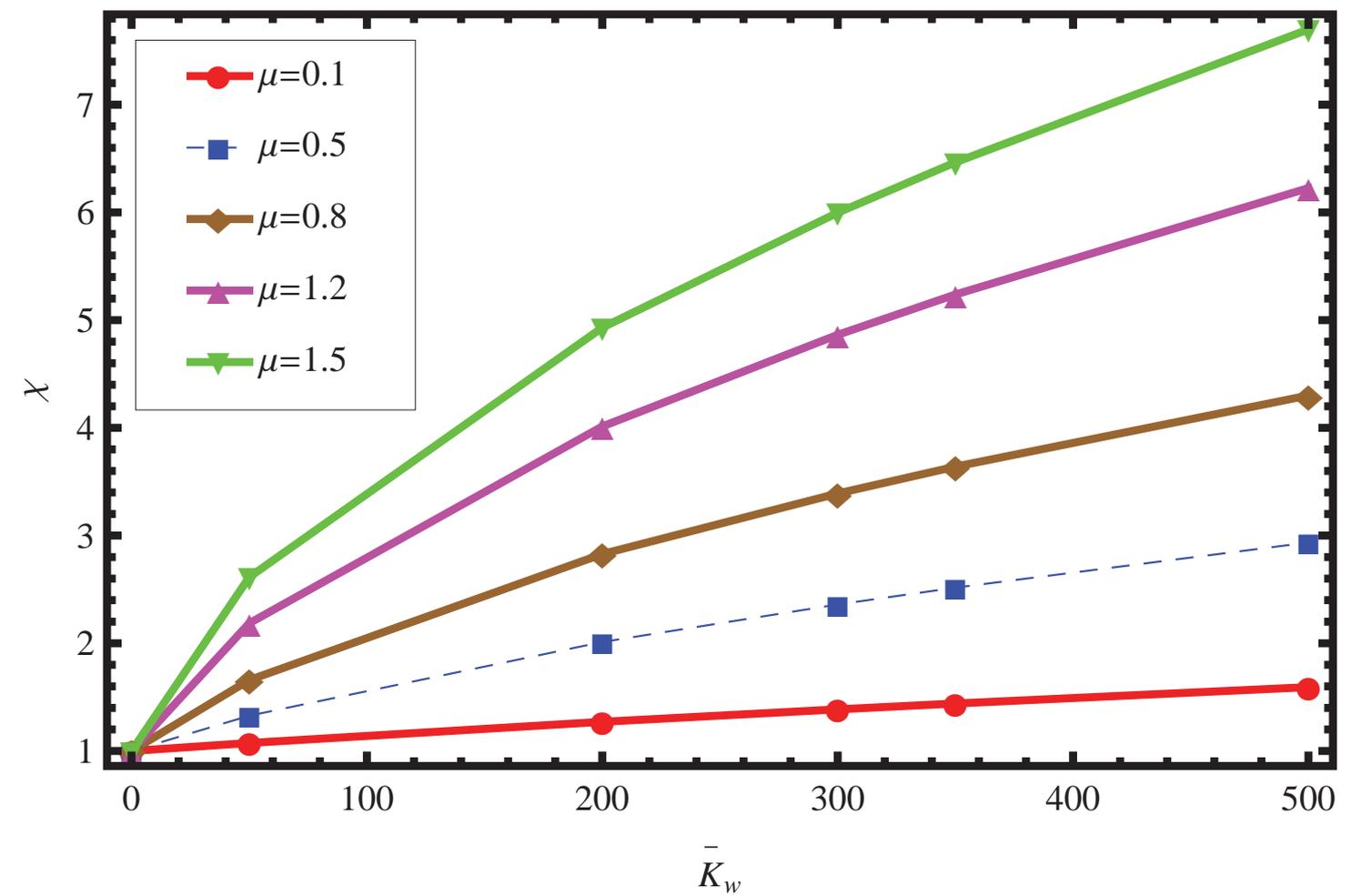

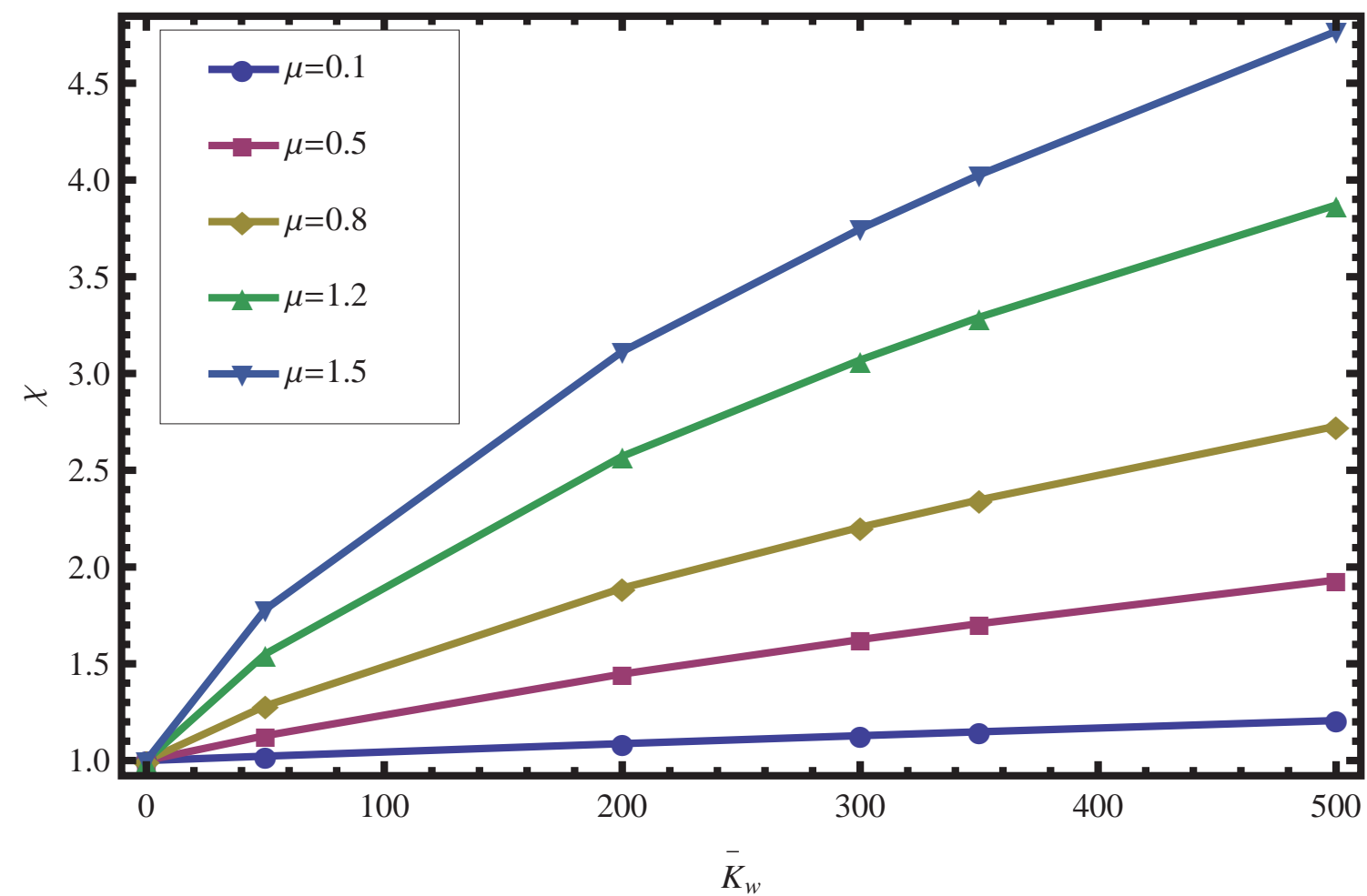

**Figure 7a**

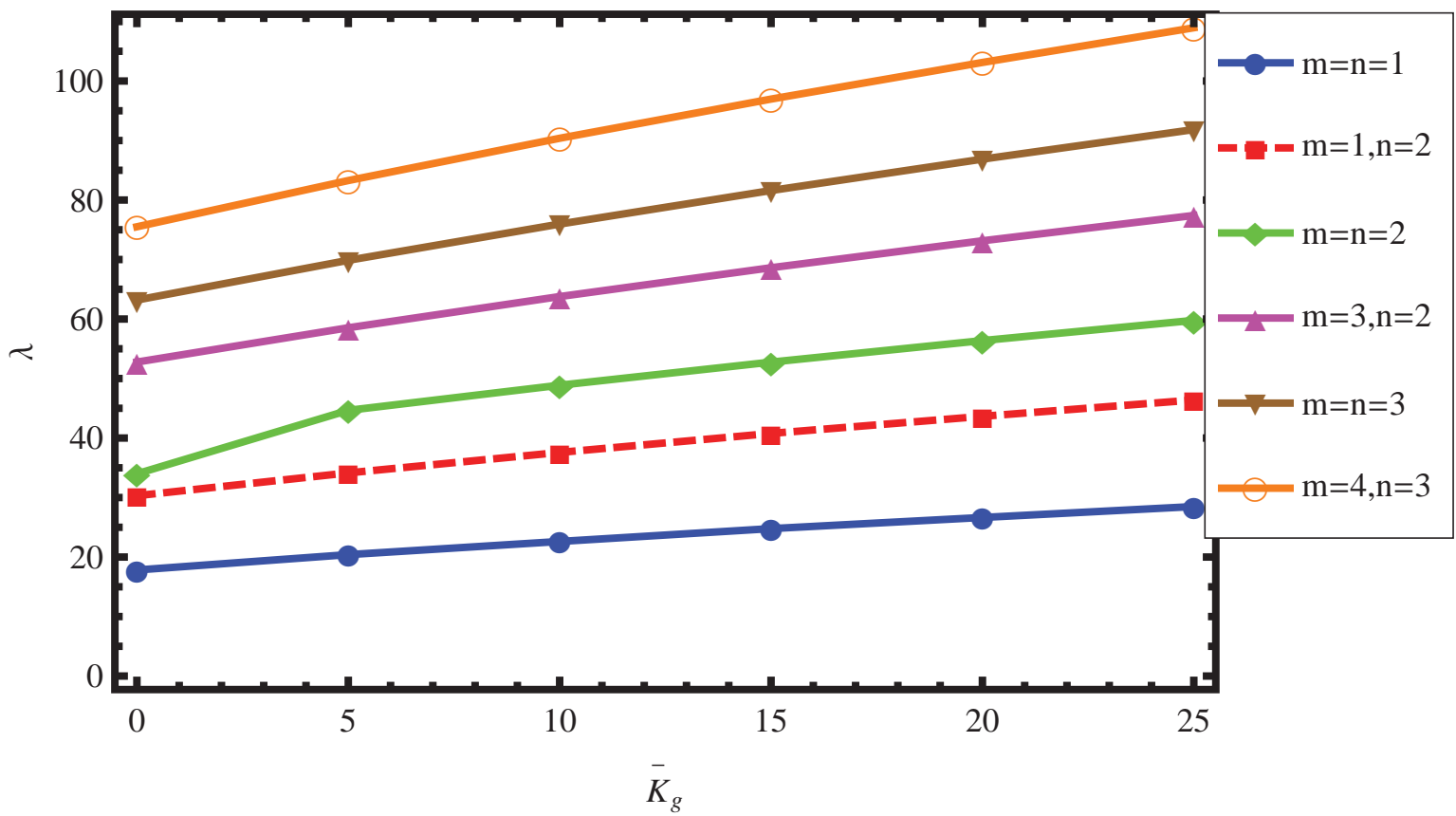

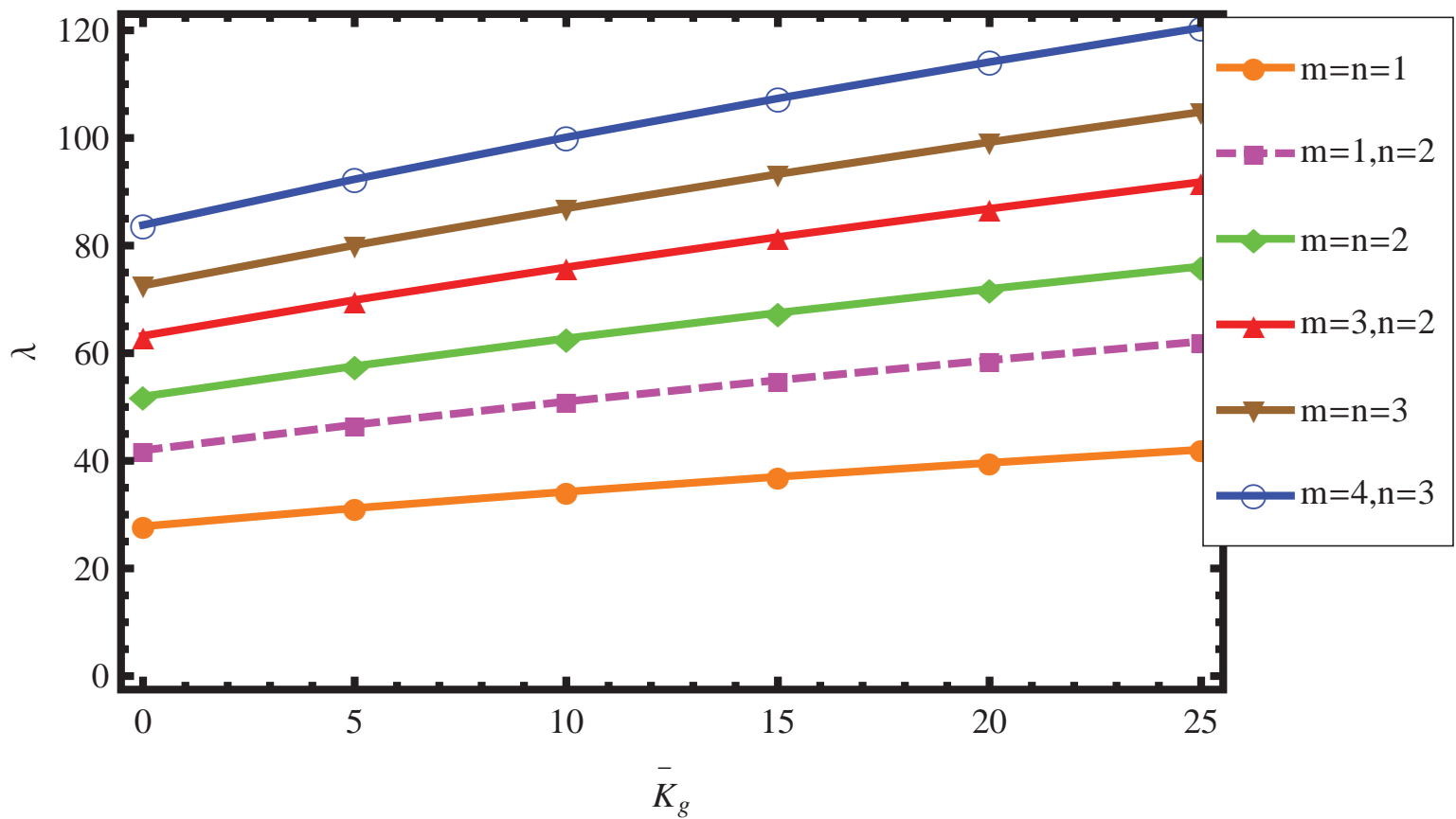

Figure 8

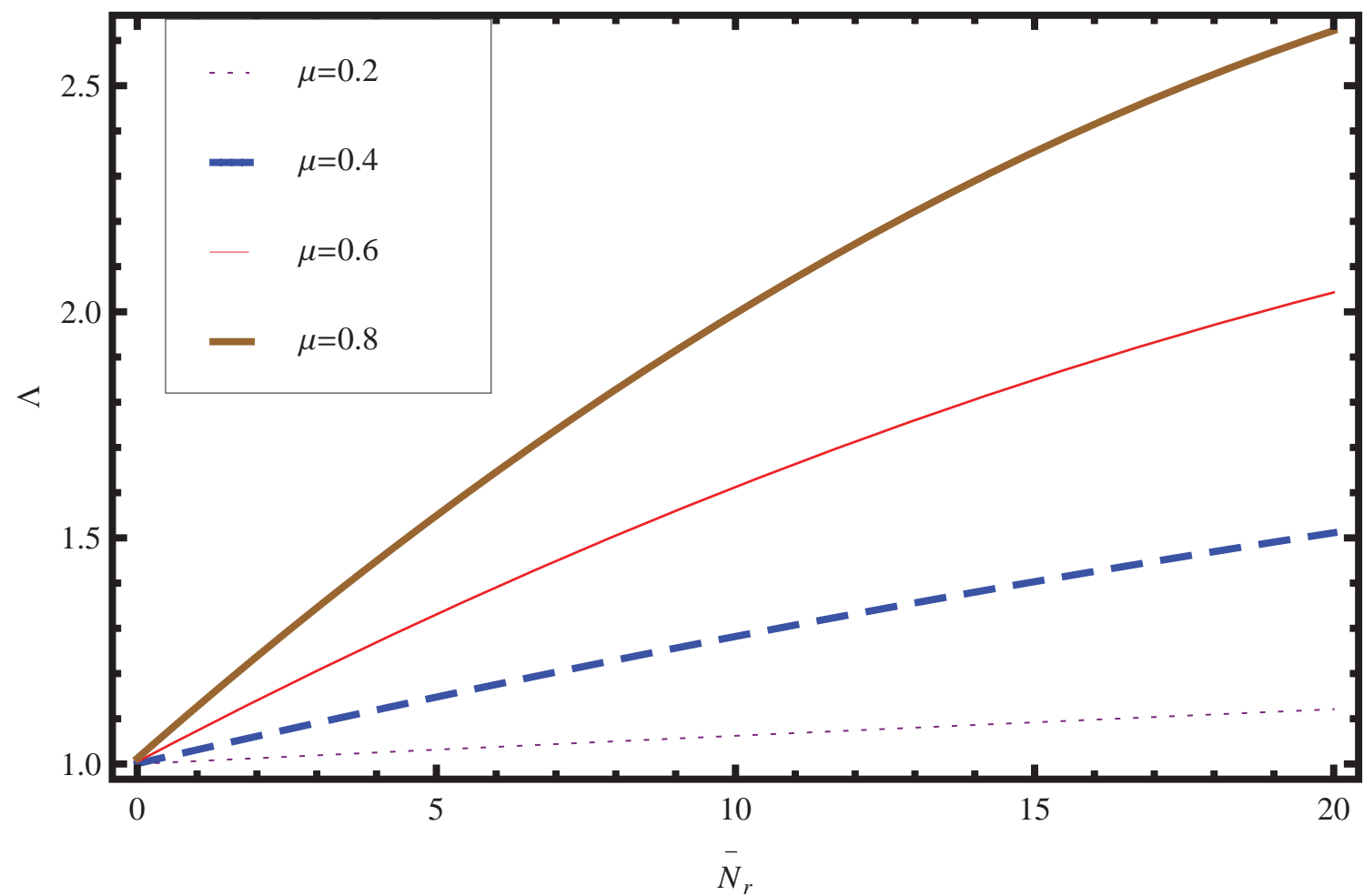

**Figure 9**

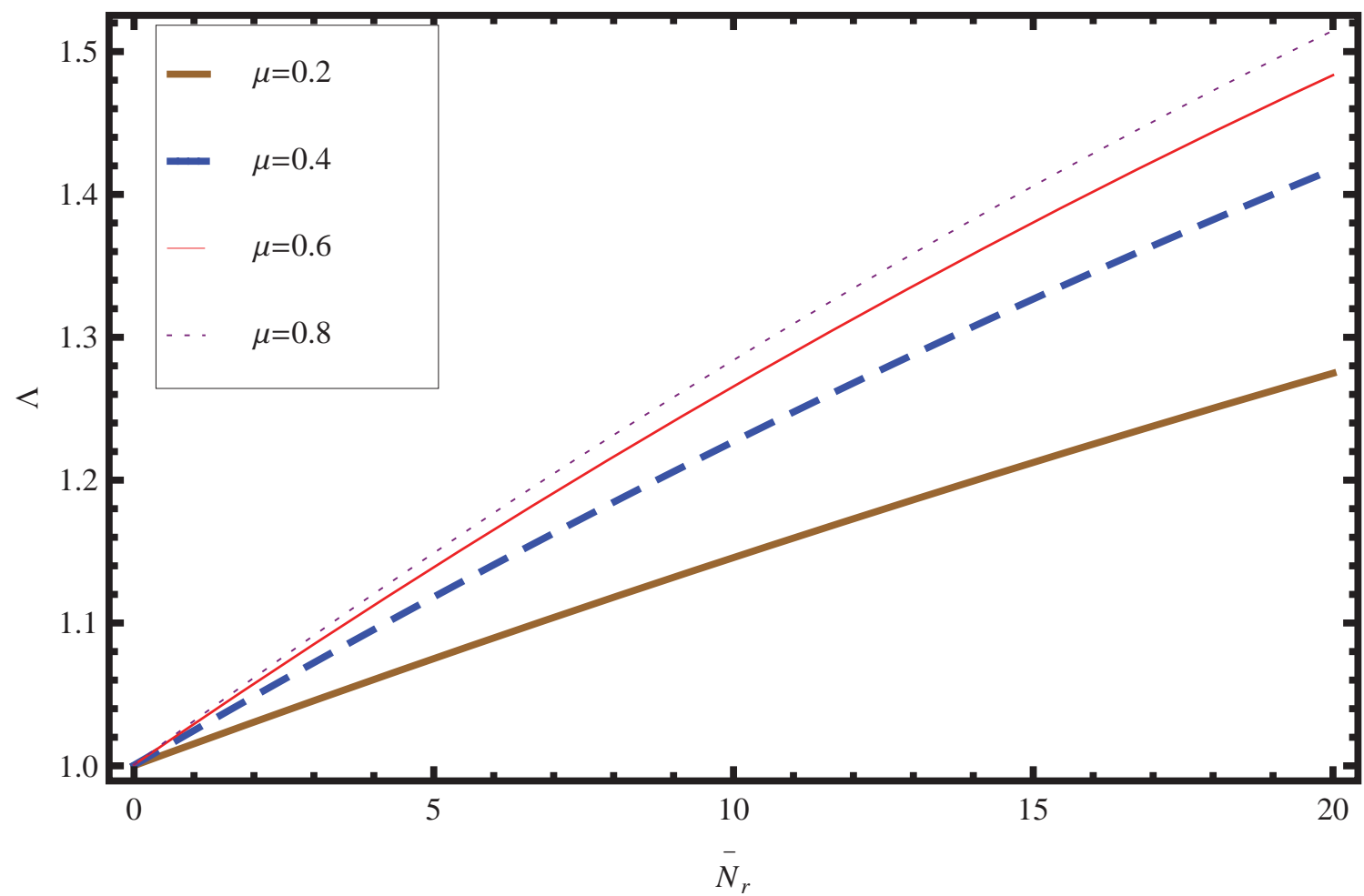

**Figures 10_11**

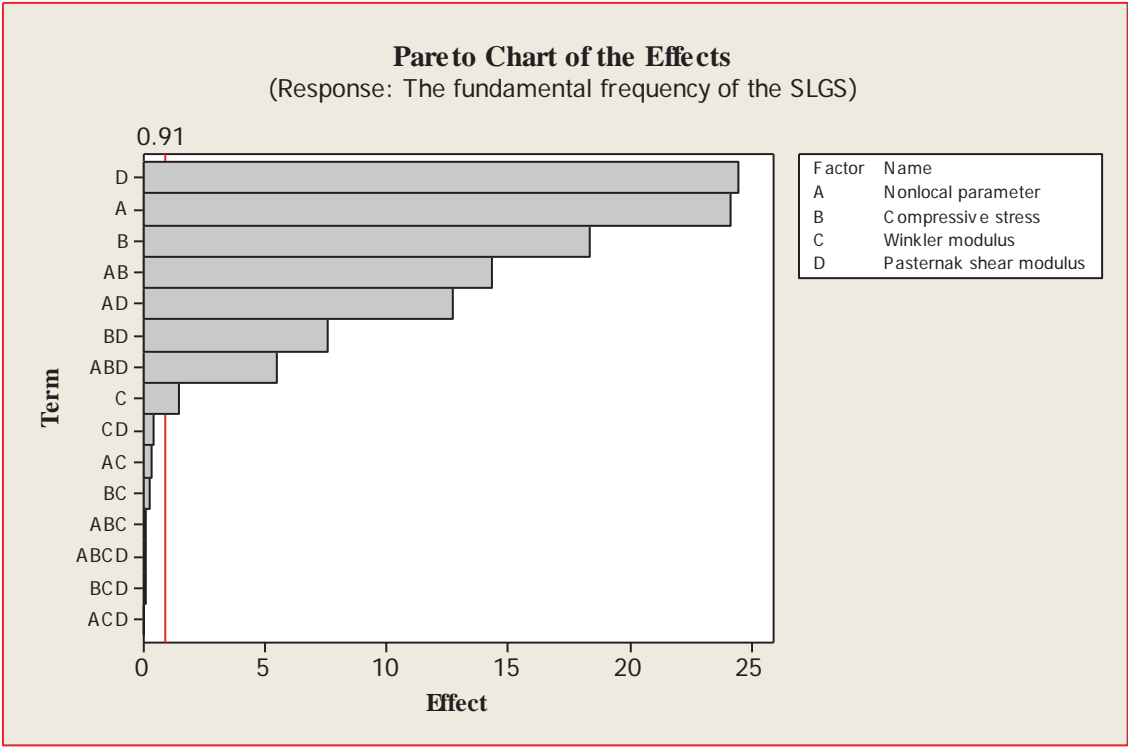

**Figure 10**. The Pareto chart of the Winkler modulus, Pasternak shear modulus, the nonlocal factor and the compressive stress for a simply-supported SLGS.

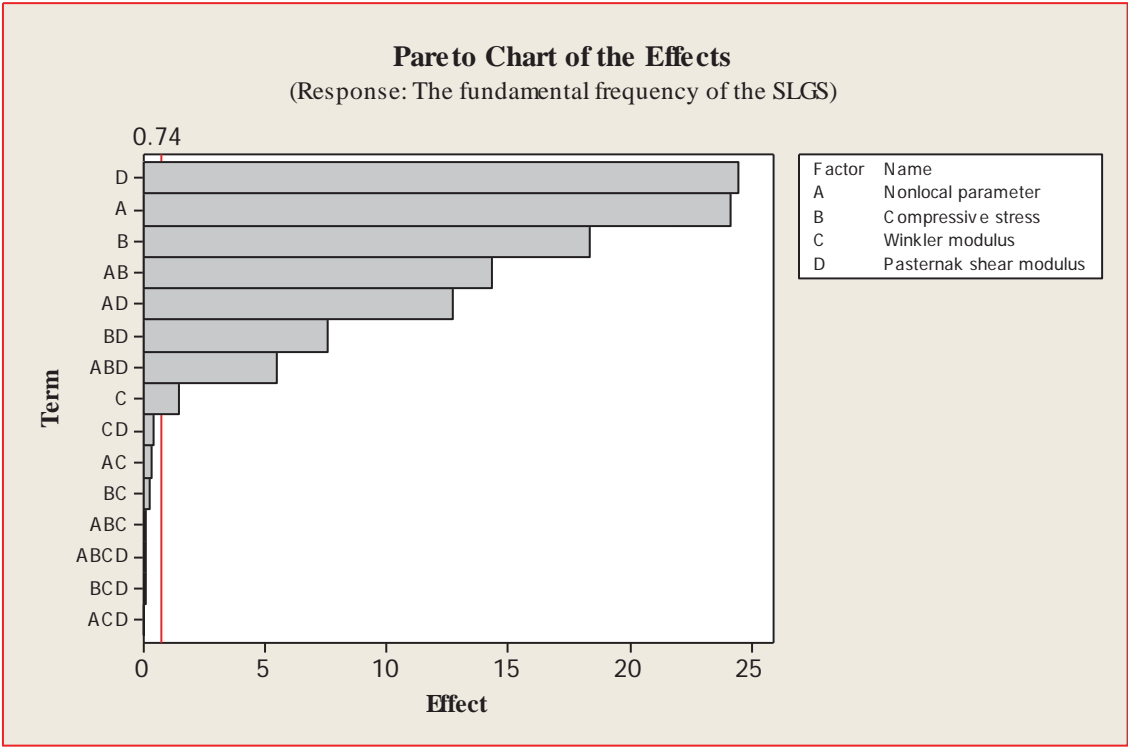

**Figure 11**. The Pareto chart of the Winkler modulus, Pasternak shear modulus, the nonlocal factor and the compressive stress for a fully-clamped SLGS.